# Water desorption and re-adsorption on epitaxial graphene studied by SPM

Tim L. Burnett, Jack Patten, Olga Kazakova*

National Physical Laboratory, Teddington, TW11 0LW, UK

**Abstract**

We demonstrate the temperature-mediated and completely reversible process of desorption–readsorption of water on a few layers of epitaxial graphene on a *4H-SiC(0001)* substrate. We show that under ambient conditions water forms solid structures on top of the second and third layers of graphene. In the case of strained or highly defective graphene domains, these features produce strongly correlated and reproducible patterns, implying importance of the underlying defects for the initial stages of water adsorption. Hydrophobicity increases with number of graphene layers. Evolution of the water layer as a function of temperature is accompanied by a significant (two-fold) change of the absolute surface potential difference between one and two layers of graphene. *In situ* observation of water evolution during heating also potentially provides a direct method for measurement of the heat adsorption on the nanoscale.

# Introduction

Graphene, a single monolayer of graphite, has attracted the interest both of the research community and industry owing to its novel physical properties and vast potential in technological applications as a likely successor of silicon in post-Moore's law devices, various biochemical sensors, ultrafast electronics, etc[1]. Epitaxial growth on SiC is one promising method for the production of the large area graphene on a semiconducting substrate[2].

For numerous industrial applications it is essential to know and control the state of the graphene surface. The presence of water is unavoidable when graphene is exposed to air for a prolonged period of time[3] and understanding its influence is crucial for understanding of both fundamental graphene properties and functionality of devices (*i.e.* wetting, catalysis, batteries, supercapacitors, chemical and biological sensors and electronic devices). Significant effort has been dedicated to both theoretical and experimental investigation of water on graphitic surfaces. In general, the hydrophobic nature of the graphene is commonly observed and revealed on the macroscale as a large contact angle, *i.e.* up to 93°, between a water droplet (~1 μl) and graphene surface, as measured by optical methods and x-ray reflectivity[4,5,6]. However, it has also been shown that few layer graphene on top of different substrates does not significantly change the wettability of such materials as gold, silicon and copper, owing to van der Waals forces dominating the surface-water interactions[7] and, as such, the few layer graphene appears hydrophilic in contradiction to bulk graphite. On the nanoscale, these interactions are even less understood. For example, adsorption of water on graphitic surfaces and formation of ice-like structures is rather commonly observed on the microscale at low temperatures[8,9,10]. Even in ambient and elevated humidity conditions water typically forms a monolayer with an ice-like structure, as, for example, was demonstrated for a thin layer of water sandwiched between mica and graphene layers[3,11,12]. The observed order is strongly dependent on the morphology of graphitic surfaces and presence of defects (surface steps, dislocations, etc.), as they offer preferential sites for the nucleation of adsorbed water to hydrophobic surfaces. Considering electronic properties, water acts as an acceptor, leading to the shift of the Fermi level in graphene devices and significantly affecting the electrical transport properties[13],[14].

Besides the production of large-areas of graphene, forthcoming industrial needs require a large-scale contactless method for testing its electrical properties. Currently, only time-consuming, complicated and expensive methods of measurements of the electrical properties of graphene are available, including patterning of devices and subsequent transport measurements. In this case, the obtained information is generalized over the whole device and not correlated with the exact morphology of graphene and presence of structural defects or local adsorbates. Electrical modes of scanning probe microscopy, *e.g.* Electrostatic Force Microscopy (EFM) and Scanning Kelvin Force Microscopy (SKPM), provide a contactless (and, therefore, simple, widely accessible and cheap) electrical mapping of epitaxial and exfoliated graphene and extracting crucial information about graphene thickness, distribution of the electrical potential and charge, work function, etc. at the microscale. While the EFM method allows mainly qualitative mapping of the surface potential[15],[16],[17],[18], the SKPM technique provides quantitative values for the work function difference, $\Phi_s = \Phi_{tip} - eV_{CPD}$, where $\Phi_s$ and $\Phi_{tip}$ are work functions of the surface and tip, respectively and $V_{CPD}$ is the contact potential difference directly measured by SKPM. Using SKPM method, both exfoliated[19] and epitaxial[20],[21],[22] graphene with different number of layers has been studied. In epitaxial graphene grown on SiC, the method has proved to be useful for easy identification of graphene domains. For example, SKPM measurements performed in vacuum exhibit a difference in the work function of ~135 meV between one and two layers of graphene (1LG and 2LG) [20]. At the same time, the properties of graphene are strongly influenced by surface charge imposed by adsorbates. Another practical aspect to consider is that to achieve large-scale and industry-relevant electrical mapping of epitaxial graphene, the measurements should be performed in ambient conditions and a practical range of temperatures. Most SKPM measurements performed in air showed reduced and significantly varying values of $\Delta\Phi_s$, see *e.g.* Ref.[19]. These variations might be attributed to different doping levels strongly dependent on sample preparation and environment. In particular, a thin layer of water can significantly affect the surface potential and lead to the reduced value of $\Delta\Phi_s$[12].

To fully exploit possibilities of graphene devices, it is essential to understand the interaction between the graphene surface and atmospheric water. In this paper we demonstrate the reversible process of the desorption–reabsorption of water on epitaxial graphene and the consequent change of the surface potential, as studied by SKPM and EFM techniques in ambient conditions and at elevated temperatures. We follow the morphological transformation of a monolayer of water on graphene surface as a function of temperature, accompanied by a significant (two-fold) increase of $\Delta\Phi_s$ as water evaporates. The results demonstrate the importance of surface studies of graphene in ambient and elevated temperature and humidity conditions, *i.e.* resembling the industrial environment, in order to fully understand the underlying phenomena and control electronic properties of proposed graphene devices.

## Results

1. Identification of layers through SKPM/EFM measurements

The sample studied here consists of nominally ~1.5 layers of graphene as revealed by Raman spectroscopy studies performed at multiple sites, *i.e.* ~60% of single and ~40% of double layers of graphene with a few individual islands of multilayer graphene (most likely triple layer), referred to in this paper as 1LG, 2LG and MLG, respectively. Identification of epitaxial graphene domains using topographical information is not straightforward due to a complex morphology of the SiC surface where terraces with a typical height of 1-10 nm form during the high-temperature annealing process. 1LG can be identified as the region of the lowest potential, forming a continuous background to the islands of 2LG. When the sample is exposed to air, decoration with 0D adsorbates may occur. The adsorbates are mainly attached to 1LG (Fig. 1a,b), whereas 2LG domains remain pristinely clean. Despite the complex morphology of the surface, the SKPM image clearly reveals two clear levels of contrast, (Fig. 1c). The surface potential images show a very clear contrast where the areas of high potential are attributed to the 2LG, and the regions of low potential correspond to the 1LG. The histogram analysis of the acquired SKPM data is shown in Fig. 1d. The histogram reveals two peaks corresponding to 1LG (left) and 2LG (right) domains and a surface potential difference of $\Delta V_{CPD}$~15 mV is measured.

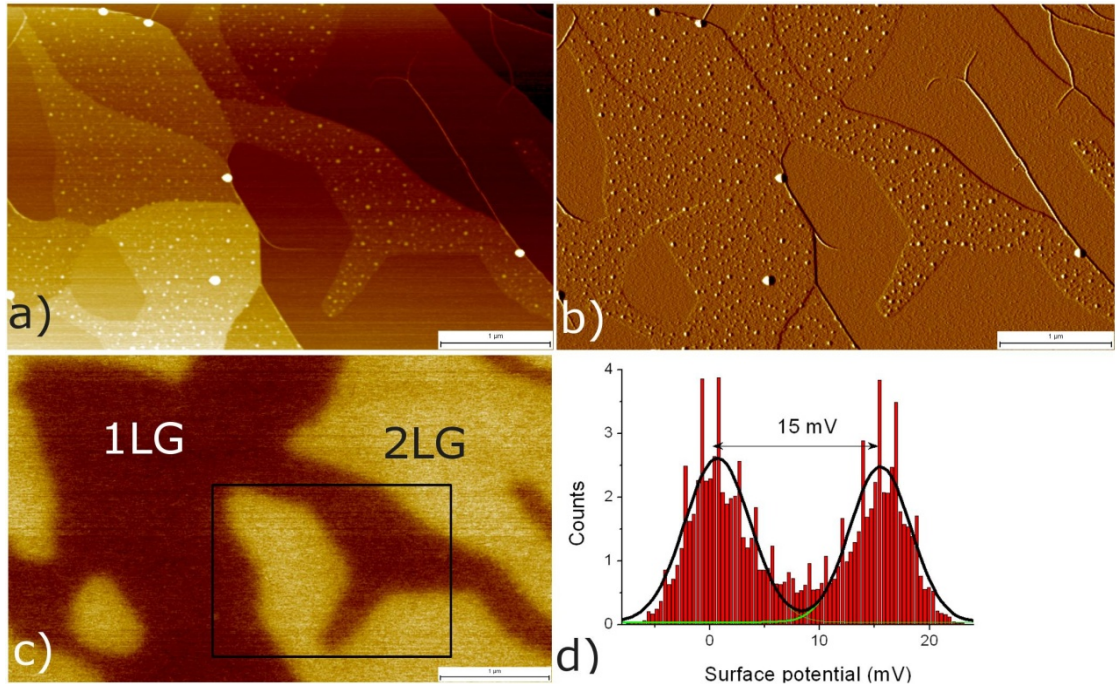

Fig. 1. Epitaxial graphene on *4H*-SiC(0001). 1LG layer is decorated with 0D adsorbates. a) Topography (z-range is 13 nm), b) topography deflection error signal highlighting the adsorbates found on the 1LG, c) SKPM image corresponding to a). The frame indicates the area where the SKPM histogram (d) was taken. The solid line in d) is a double-peak fit using Lorentzian distribution. Measurements performed at room temperature. The scale bar in a-c) panels is 1 μm.

2. SKPM at elevated temperatures

As the measurements were performed in air (relative humidity ~40%), atmospheric molecules are likely to adsorb on the sample surface[14], affecting the overall surface potential, $\Delta V_{CPD}$ of the graphene surface. To explore this, we performed sequential heating of the sample, *i.e.* T = 20, 50, 80 and 120* °C (see Methods). Topography, tapping phase and potential images were obtained at a fixed temperature after leaving the sample to equilibrate at the set temperature for one hour. The same area was monitored through this series of temperatures. The area shown in Fig. 2 contains two distinct domains, further referred as 'heart' and 'stripe' (see the frame in Fig. 2b). It should be noted that no notable decoration with 0D adsorbates occurred in the chosen area (Fig. 2a). However, by comparison with results shown in Fig. 1, the intermediate bright areas (heart and the bottom left part of the stripe) in Fig. 2c were identified as 2LG. Additionally, the brightest contrast (the top right part of the stripe) reveals this area as a MLG domain. The corresponding EFM image of the same area (Fig. 2d) taken with the tip bias -1.5 V shows the change of the cantilever phase due to the change of the capacitance between the tip and substrate. It should be noted that the EFM image provides significantly better sensitivity than the SKPM method, giving improved spatial resolution for the same lift height.

Fig. 3a shows linear profiles of the potential images obtained at 20 and 120* °C. The initial $\Delta V_{CPD}$ between 1LG and 2LG is ~33 mV and between 2LG and MLG is ~22 mV at 20 °C. As the temperature was increased to 120* °C, a significant increase of the surface potential difference between 1LG and 2LG was observed, *i.e.* $\Delta V_{CPD} \cong 52$ mV. However, no notable change in the $\Delta V_{CPD}$ value between 2LG and MLG occurred at 120 °C (Fig. 3a).

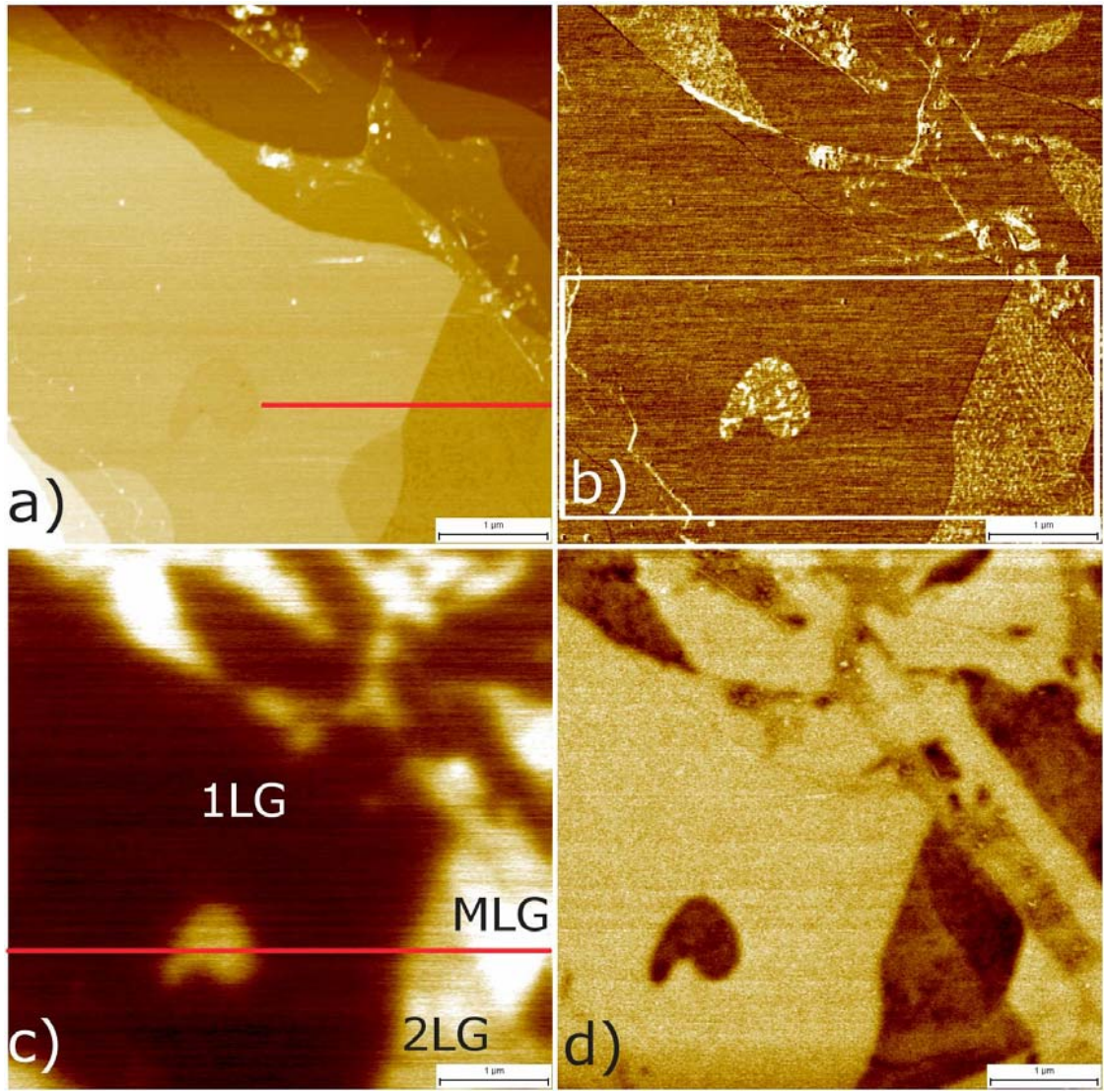

Fig. 2. Area of interest typical of the studied graphene sample at room temperature: a) topography; b) tapping phase. Phase image clearly shows an inner pattern on 2LG and MLG suggesting inhomogeneity in the structure of the graphene; c) SKPM potential and d) EFM phase image, $V_{tip}$ = -1.5 V. Profiles taken along the red lines in a) and c) are analysed in Fig. 5 and 3a, respectively. The scale bar is 1 μm.

Furthermore, we performed sequential heating-cooling measurements recording the surface potential difference, initially heating the sample from 20 to 120* °C and then cooling it back to room temperature in the same number of steps. At each temperature the potential image of the same area (as presented in Fig. 2 b) was obtained, and the heart and stripe domains with surrounding 1LG were analysed using the histogram method. A summary of the results is presented in Fig. 3b. For the 1LG-2LG area the $\Delta V_{CPD}$ steadily increases with temperature, reaching ~50 mV at 120* °C. During the cooling process, the $\Delta V_{CPD}$ values gradually reduce, remaining, however, generally higher than during heating. Once the sample was cooled back down to room temperature, $\Delta V_{CPD}$ also returned to its initial value of ~26 mV. Despite returning to the same value, the temperature dependence of $\Delta V_{CPD}$ between 1LG and 2LG demonstrates a clear hysteresis. Different behaviour was observed for 2LG-MLG area: first $\Delta V_{CPD}$ sharply increases at 50 °C from 6 to 16 mV, as temperature increases further $\Delta V_{CPD}$ drops back to the initial value of ~6 mV. Hysteretic behaviour in this case is significantly less pronounced. A convolution of these two responses is reflected in the change of $\Delta V_{CPD}$ for 1LG-MLG area, where the surface potential difference quickly increases at 50 °C and remains at ~60 mV irrespective of further heating. Overall, a significant, nearly two-fold increase of $\Delta V_{CPD}$ for 1LG-MLG domains is observed upon heating.

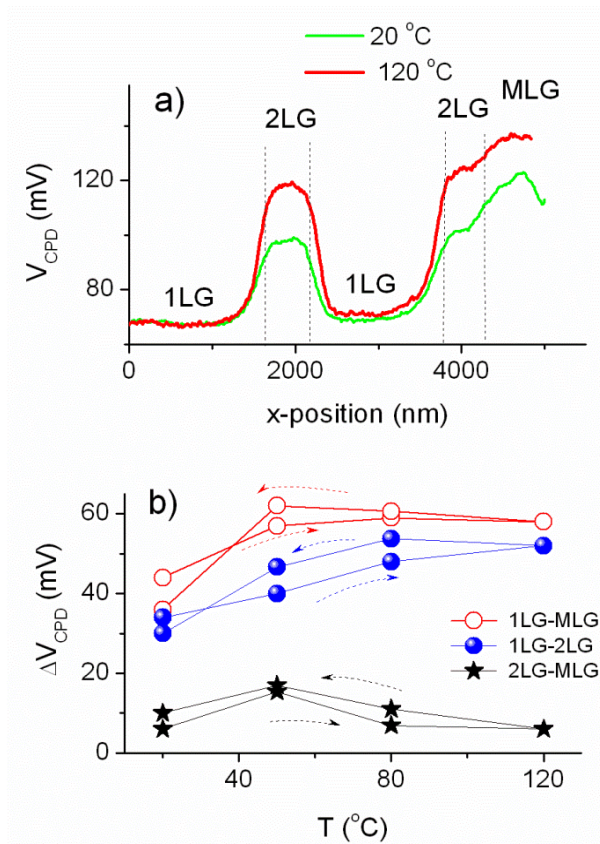

Fig. 3. a) Linear profiles of the surface potential difference at 20 and 120* °C along the line shown in Fig. 2c. Vertical dashed lines show the approximate boundaries of 2LG domains. b) Temperature dependence of the change of the surface

potential difference obtained using histogram analyses of subsequent potential images for 1LG-2LG, 2LG-MLG and 1LG-MLG areas. The arrows show directions of the temperature change. The solid lines are guides for the eye.

3. Tapping phase at elevated temperatures

The tapping phase images demonstrate the uniform and homogeneous nature of the 1LG domain for the full temperature range (Fig. 4). However, the phase images of 2LG and MLG domains are significantly different: at room temperature a substructure can clearly be seen, which in the case of the MLG domain forms a well-organized pattern (Fig. 4, RT1). It should be noted that such substructure is not observed in potential images, most probably due to a relatively low spatial resolution of the method, ~50 nm. This is confirmed by the fact that better spatial resolution afforded by the EFM image allows some substructure to be seen (Fig. 2d). As the temperature increases, the nature of the tapping phase image changes considerably. Already by 50 °C the substructure within the MLG domain is significantly modified and can be characterized by larger features. At 80 °C, the substructure disappears completely and both 2LG and MLG domains become almost entirely uniform. As the temperature returns back to 20 °C, the pattern reappears on both 2LG and MLG (Fig. 4, RT2). Moreover, the same type of pattern consisting of well-arranged parallel lines can be clearly seen on MLG, demonstrating a complete reversibility of the process. It should be noted that even at the highest temperature the resolution of the tapping phase and topography does not deteriorate, as can be seen from the many features, *i.e.* the boundary between 1LG and 2LG regions, can still be clearly resolved. Whilst the phase contrast for the 1LG domain is generally temperature independent, with a standard deviation from the mean value being ~0.25°, for 2LG this parameter changes significantly with temperature, being in the range 0.32-0.66° (see Table I). This analysis implies the presence of an additional layer of a different material on top of graphene domains and is reflected by its reversible nature with respect to temperature and the contrast observed in the tapping phase images. Comparison of topography profiles taken at RT1, 80 °C and RT2 (where RT1 and RT2 are initial and final measurements at room temperature, respectively) across 1LG and 2LG domains (separated by a relatively high terrace step) is shown in Fig. 5. The height of the additional layer is consistently ~400 pm as measured on 2LG at room temperature. However, it should be noted that due to the minute lateral dimensions of the layer and absence of well-defined plateaus the accurate measurements of the height are rather restricted. Further analyses of topography images is summarised in Table I. At RT1 the 2LG is characterised by a greater roughness. The roughness of both 1LG and 2LG domains decreases significantly and becomes almost identical at 80 °C. Upon the following cooling down to RT2, the initial roughness of the 2LG restores.

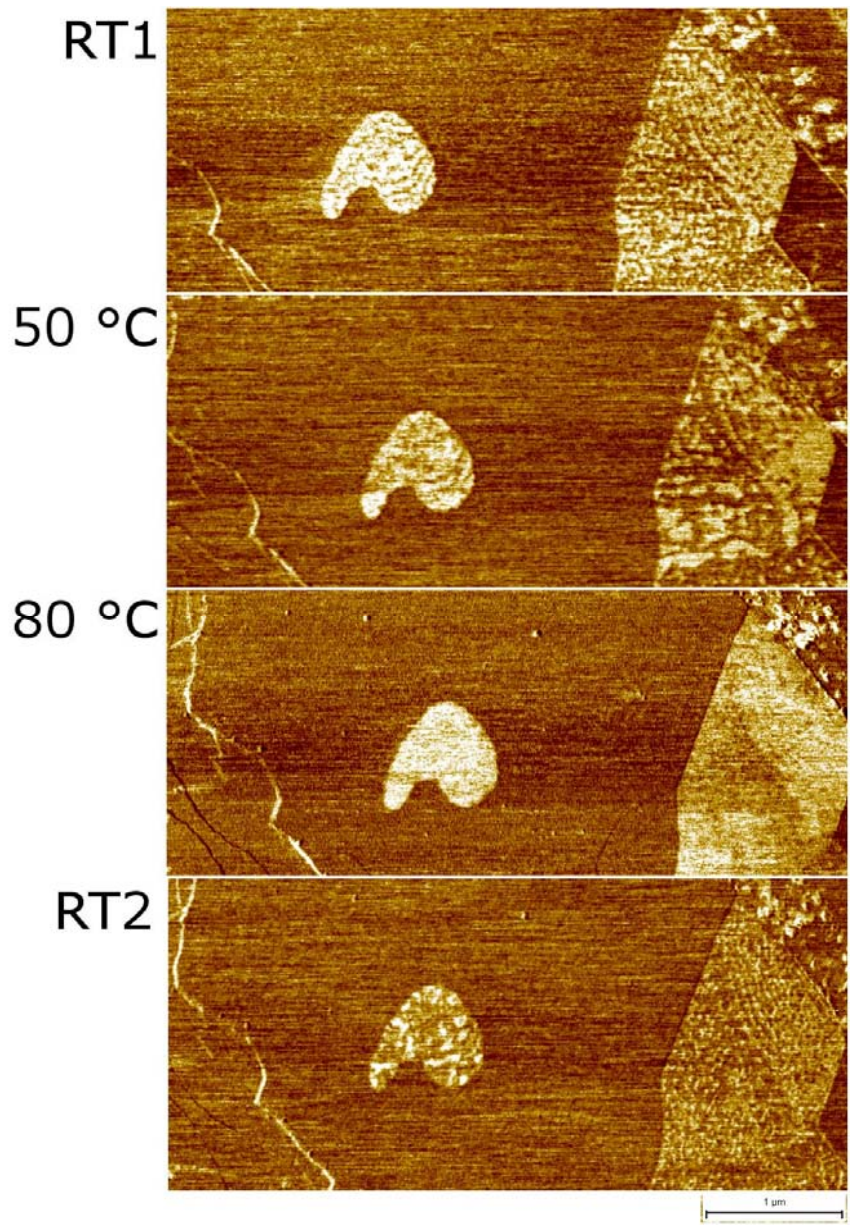

Fig. 4. Tapping phase images of heart and stripe domains corresponding to the frame in Fig. 2b and obtained at different temperatures, top to bottom: RT1, 50 °C, 80 °C and RT2.

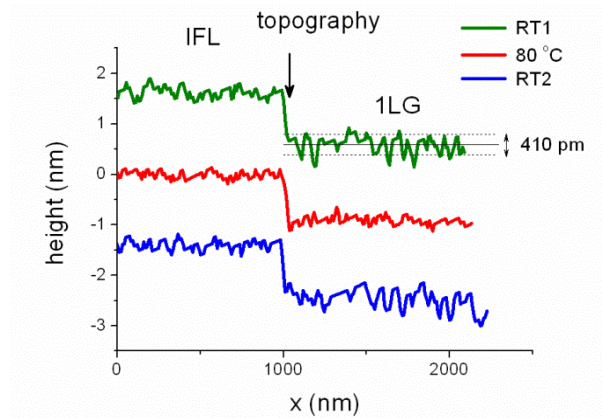

Fig. 5. Topography profile of the 1LG-2LG domains along the line shown in Fig. 2a obtained at RT1, 80 °C and RT2. Profiles are vertically offset for clarity. The vertical arrow indicates the position of the SiC terrace of ~1 nm height dividing the 1LG and 2LG domains. The horizontal solid line shows the mean value for 2LG at RT1 and the two dashed lines define the spatial resolution of the method, corresponding to 2σ-value of the Gaussian distribution.

Table I. Standard deviation from the mean value for the tapping phase and topography. RT1 and RT2 indicate initial and final measurements at room temperature, respectively.

| Temperature | Standard deviation | | | |
|---|---|---|---|---|
| | Tapping phase (deg) | | Topography (pm) | |
| | 1LG | 2LG | 1LG | 2LG |
| RT1 | 0.27 | 0.48 | 111 | 168 |
| 80 °C | 0.29 | 0.32 | 73 | 76 |
| RT2 | 0.21 | 0.66 | 150 | 291 |

**Discussion**

Scanning probe microscopy techniques have been widely used for studies of thin water layers on various substrates in ambient conditions. While the AFM tip may strongly perturb the nanoscale liquid film, the use of non-contact modes of scanning was shown to significantly reduce the friction and dragging effects. It was generally found that initially a continuous ice-like layer of structured water is formed even at ambient temperature and humidity on both hydrophilic and hydrophobic substrates, such as mica, $BaF_2$, quartz, graphite, self-assembled monolayer, etc. (see *e.g*. Ref. 23). This ordered ice-like structure can be characterised by the lack of free OH groups. These examples demonstrate that water can form ordered, solid structures in ambient conditions even on essentially hydrophobic surfaces. Moreover, the AFM technique in general (taking necessary precaution conditions) is suitable for successful visualization and investigation of such structures.

Based on above considerations and by comparison of temperature dependencies of both SKPM and tapping phase images presented in this paper, the substructures appearing on 2LG and MLG domains are attributed to the presence of a monolayer of water on epitaxial graphene. Topographical images of these substructures in the case of epitaxial graphene closely resemble morphology of water as shown in the previous studies[3,11,12]. The height of the features corresponds to the height of monolayer water, *i.e.* ~400 pm (Fig. 5), and the area coverage is consistent with relative humidity of 40%[3]. Tapping phase images show contrast within these substructures (Fig. 4), suggesting that the observed features are located on top of the graphene.

We believe the formation of well-organized pattern of water at room temperature is facilitated by defects and likely to reflect distribution of defects and/or strain in 2LG and MLG domains[24,25]. Inhomogeneity of the substrate affects the morphology of the graphene and is likely to cause changes in the surface corrugation of the graphene, this could in turn mediate alterations in substrate-related doping. The resulting variations in charge or surface geometry might be considered as dominating mechanisms which promote or demote adsorption of water[26] and as such define the properties of different layer thicknesses of graphene. Therefore, combination of purely geometrical (defects, corrugation) as well as electrostatic (dangling surface bonds and associated dipole charges) origins can govern water adsorption to epitaxial graphene. These considerations help to answer an important question arising from our studies, which is: why within the same sample of epitaxial graphene we observe regions which appear nearly completely covered with water (see Fig. 1 in Supplementary Information) and other areas which have no visible water (*i.e.* Fig. 1)? For example, the heart area, evidently being a topographically isolated 2LG island, is expected to be in a more relaxed state than the stripe area. Correspondingly, little ordering in the pattern of water is observed within the heart area, whereas the stripe area shows a clear well-ordered pattern of adsorbed water likely decorating areas of the high defect concentration possibly induced by the surrounding terraces and debris of the graphene growth.

Furthermore, when the sample is subjected to the step-wise increase in temperature, the substructures under study change and ultimately disappear at around 100 °C (Fig. 4). On cooling the sample back down to room temperature, the same type of substructure can be observed, although with a slightly different exact configuration. This implies that the water initially desorbs from the surface and then returns to the preferential adsorption sites upon cooling but this process is not exact as would be expected. This behaviour further supports our initial assumption of the presence of monolayer water on epitaxial graphene and its desorption/reabsorption as temperature increases/decreases. Disappearance of water at elevated temperatures also proves that it was initially resided on top of the graphene layer due to condensation of water vapour from the atmosphere, in contrast to the case of exfoliated graphene where water was trapped between the layer of graphene and the substrate[3,11,12]. Furthermore, very good reproducibility of the experimental results was observed in numerous experiments on thermal cycling followed by SKPM and tapping phase imaging. In particular, the same temperature-related modification of the pattern reflected desorption/reabsorption monolayer of water was observed after several months after the initial experiments on the same sample.

Additional experiments made on exfoliated graphene on $SiO_2$ at room temperature (Fig. 2, Supplementary Information) demonstrate that adsorbed water can also be observed only on the 2LG and 3LG domains, *i.e.* similar to the situation seen for the epitaxial graphene. A characteristic angle of ~120° was also frequently observed. These water islands evolve during the measurement without changing the external conditions. In the case of exfoliated graphene, the coverage and morphology were consistent with a monolayer water coverage investigated in the present studies of epitaxial graphene and those reported in the literature[3,11,12]. Due to significant roughness of the $SiO_2$ substrate, ≤500 pm, it was not feasible to directly measure the height of the monolayer of water, which was within the same height range.

Although hydrophobic properties of graphene and other graphitic surfaces are commonly accepted, and the general difficulty in visualization of water on such interfaces was discussed above, in this work we have succeeded in mapping of a monolayer of water on graphene domains in ambient conditions. We believe that the reason for this is the formation of solid ordered structures of water on the nanoscale, which provide a stable surface for imaging. It is well-known that freshly prepared pristine SiC is strongly hydrophilic, with water contact angle for a freshly prepared Si-side surface being only ~15°[27]. When carbon bonds of any nature appear on top of SiC (including formation of an interfacial C-rich layer), the hydrophilic properties get less pronounced and the contact angle may increase up to ~85°[27,28], though still remaining significantly less hydrophobic than in multi-layer graphene and graphite. However, as has been recently demonstrated for silicon, gold and glass substrates, few layers of graphene do not dramatically change wettability of these hydrophilic substrates, where, for example, the contact angle changes from 20.2 to 54.3° in the case of pristine glass and 2LG/glass, respectively[7], still remaining significantly less than the expected, ~93°[4,5,6]. In the case of epitaxial graphene, the wetting properties of a thin graphene layer are dominated by SiC due to their strong interaction, which essentially results in the fact that we have a more hydrophilic graphene than that predicted. These results also point out the importance of considering graphene properties only in conjunction with those of the substrate, which not only affect the charge transport and carrier type and concentration but also may lead to modification of the chemical properties of graphene, and as shown here, its affinity to water.

The observed morphological changes are fully consistent with a gradual evolution of the relative surface potential difference of the graphene with temperature (Fig. 3). The change in $V_{CPD}$ is reversible in the same way that was observed in the tapping phase images. Additionally, the temperature dependence of the surface potential has a clear hysteresis with the width of ~10-15 mV at 50 °C, reflecting the fact that the desorption process was generally slower than the reabsorption in the intermediate temperature range, probably originating from the history of the sample, *i.e.* longer exposure to humidity. We also observe minor differences in how the monolayer of water evolves as a function of temperature on 2LG and MLG, however see no changes in the 1LG (Fig. 3). Our nanoscale observations support the proposition that hydrophobicity increases with increasing layer thickness, such that the hydrophobicity of SiC<1LG<2LG<MLG. This also agrees well with the previous macroscopic studies and results of molecular dynamic modelling of water on epitaxial graphene. Our experimental results can be understood such that 120 °C does not provide sufficient energy to desorb the water from the 1LG and so its appearance remains unchanged. The observation that the water desorbs most quickly from the MLG is consistent with the fact that MLG is more hydrophobic than 2LG, so less heat of desorption is required to achieve this. These considerations are generally consistent with a value of the adsorption energy for water molecules being significantly larger on pure SiC (-636 meV)[29] than on graphene (18-47 meV)[30]. We also believe that *in situ* observation of water evolution at this scale during heating can potentially provide a direct method for measurements of the heat adsorption on the nanoscale.

## Summary

We present studies of controlled, fully reversible and temperature-dependent process of desorption and reabsorption of water molecules on top of a few layers of epitaxial graphene domains in the temperature range 20 -120 °C. In this work we show direct measurements of the surface water on graphene for two reasons: 1) water monolayers appear to be relatively immobile at room temperature With nucleation of solid structures is dominated by the presence of defects and strain in the graphene domains and takes form of a clear, well-ordered and stable pattern; 2) SiC is hydrophilic and, as has already been reported for few layer graphene on top of hydrophilic substrates, the surface interaction is dominated by the substrate, again providing surface water that is sufficiently stable to image with AFM.

With increasing temperature, water molecules desorb from graphene, leading to the disappearance of the characteristic pattern and significant (up to two times) increase of the surface potential. We note that the monolayer of water desorbs initially (*i.e.* at ~50 °C) from MLG and only later (~100 °C) from 2LG, showing higher hydrophobicity (*i.e.* lower binding energy) between graphene and water of the MLG compared to 2LG. However, as the temperature returns back to 20 °C, readsorption of water on 2LG and MLG domains was observed, as demonstrated by tapping phase images and the corresponding decrease of the surface potential back to its initial room temperature value. Overall, these nanoscale experiments reveal lower than expected hydrophobicity of the graphene. Thin graphene layers only moderately modify the intrinsic wettability of the SiC substrate, allowing for formation of solid water structures typically observed on hydrophilic surfaces. Thus, it is important to stress that making assumptions on the relative hydrophobicity or hydrophilicity of graphene surfaces should be treated with caution on the nanoscale and such essential aspects, as production method, type of substrate and number of graphene layers must be taken into account.

The results demonstrate significance of surface studies of graphene in ambient and elevated temperature conditions in order to fully understand the underlying phenomena and control electronic and surface properties of future devices. Imaging of water surfaces with nanometer resolution and correlation of the presence of water with the surface potential distribution of graphene allow for significant insight into the control of wetting process of graphene, essential for its applications in electronics and sensing devices as well as in conformal and impermeable surface coatings.

**Method**

The submonolayer epitaxial graphene was prepared by sublimation of SiC and subsequent graphene formation on the Si-terminated face of a nominally on-axis *4H-*SiC(0001) substrate at 2000 °C and 1 bar argon gas pressure. Details of the fabrication and structural characterization are reported elsewhere[31]. The specific synthesis route has been developed to provide large areas of homogeneous graphene layers[2,32]. The resulting material is n-doped, owing to charge transfer from SiC, with the measured electron concentration in the range 5-10×$10^{11}$ (6-20×$10^{11}$) cm$^{-2}$ and mobility of ~4000-7500 (~3000) cm$^2$ V$^{-1}$s$^{-1}$ at low (room) temperatures[33],[34].

The measurements were conducted on a Bruker Icon AFM. Bruker SCM-PIT Pt-Ir coated probes with a tip radius of ~10 nm and a force constant of ~3 N/m were used in the experiments. Topography height images of epitaxial graphene have been recorded simultaneously with tapping phase and surface potential difference maps either using amplitude modulated Scanning Kelvin Probe Microscopy (SKPM) or Electrostatic Force Microscopy (EFM) phase shift.

In tapping phase mode the phase lag between the cantilever oscillation and the original signal is monitored. The phase lag is very sensitive to mechanical inhomogeneity of the sample and variations in its material properties, allowing detection of deviations in mechanical properties and composition of the material, including modulus, adhesion, friction, viscoelasticity, etc.

SKPM is a two-pass technique that maps the electrostatic potential, $V_{dc}$, on the sample surface at a certain lift height[35]. In the first pass, the topography and the tapping phase contrast are acquired. During the second pass, the cantilever is excited electrically by applying an ac voltage to the tip. The cantilever experiences a force when the potential on the surface is different from the one on the tip. SKPM is a nulling technique, *i.e*. the dc tip potential is being changed by varying the voltage of the tip until the average force experienced due to the ac component vanishes, and thus the potential on the tip is at the same potential as the region of the sample directly underneath it. Mapping $V_{dc}(x)$ reflects distribution of the surface potential $V_{CPD}(x)$ and correspondently the work function $\Phi_s$ along the sample surface. For the SKPM measurements a lift height of 25 nm and an ac voltage of 2 V at 500 Hz were used.

Amplitude modulation EFM is also a two-pass method where the electrostatic force is measured between the charge on the sample and the tip (approximately a point dipole). This force is detected as a change in the resonant frequency, recorded as a phase shift of the resonance of the cantilever as it experiences either an attractive or a repulsive force[36]. For EFM experiments at room temperature a lift height of 25 nm and a tip bias in the range ±2 V were used. At elevated temperatures, however, adjustments often had to be made to maintained a good quality of the image

All images were collected in air at temperature $T$=20-120 °C. For the heating experiments the graphene sample was initially placed on a large hot plate where it was kept at a fixed temperature for one hour. The sample then was transferred inside AFM where it was placed on a small hot plate. The chosen temperature was maintained during the imaging process. This procedure allowed for efficient evaporation of water from the sample surface and at the same time prevented condensation of water vapours on the cantilever, preserving its mechanical and electrical properties. Imaging was performed at the following sequence of temperatures: 20, 50, 80, 120*, 80, 50 and 20 °C. At 120* the measurement temperature was 80 °C immediately after heating the sample to 120 °C.

**Acknowledgements**

This work has been funded by the EU FP7 Project "ConceptGraphene". We are grateful to Rositza Yakimova for providing the sample and Andy Wain, Ruth Pearce and Alexander Tzalenchuk for useful discussions.

Author contribution statement

All authors were involved in measurements, OK and TLB wrote the manuscript, all authors reviewed the manuscript.

Competing financial interests

The authors declare no competing financial interests.